\documentclass[conference]{IEEEtran}
\usepackage{amsmath}
\usepackage{algorithmic}
\usepackage{graphicx}
\usepackage{subcaption}
\newtheorem{definition}{Definition}

\begin{document}
\linespread{1.2}
\title{Incremental Information Gain Mining Of Temporal Relational Streams}


\author{
    \IEEEauthorblockN{
        Author name(s) are removed to meet the double blind submission requirement.
    }
}

\maketitle
\thispagestyle{plain}
\pagestyle{plain}

\begin{abstract}
    This paper studies the problem of mining for data values with high information gain in relational tables.  High information gain can help data analysts and secondary data mining algorithms gain insights into strong statistical dependencies and causality relationship between key metrics.  In this paper, we will study the problem of high information gain identification for scenarios involving temporal relations where new records are added continuously to the relations.  We show that information gain can be efficiently maintained in an incremental fashion, making it possible to monitor continuously high information gain values.
\end{abstract}

\IEEEpeerreviewmaketitle

\section{Introduction}

Data science has become a major component of the modern data-driven decision making processes.  The success of data science owes itself to analytical techniques and algorithms that extract hidden patterns and insights from large volumes of data.  While the precision and expressive power of the algorithms are an important part of the data science innovation, another crucial aspect of a successful data science algorithm is its efficiency and robustness.  Many data science algorithms are powerful and can detect complex patterns, they can be prohibitively expensive and inflexible for many real-world applications.  For example, while deep learning algorithms are incredibly intelligent, it is well known that they require very large volumes of high quality data trained on expensive processors (often requiring multiple high end GPU processors).  It's also known that machine learning algorithms do not generalize to data streams as well as static data sets \cite{krawczyk2017ensemble}.

In this paper, we focus on the other end of the spectrum of data science algorithms that have high efficiency and can cope with long lasting continuous relational data streams.  The algorithm we present in this paper is {\em information gain mining}.  It detects relational values that exhibit high information gain with respect to the distribution of some other attributes.  A value with high information gain highlights an exceptional degree of correlation or causal relationships to some relational attribute, and as a result, such values are of particular interest to data analysts or secondary learning algorithms to uncover hidden patterns in the data.

We study the information gain mining problem for relational data streams where the relational data is continuously updated with batches of new records.  In particular, we require the mining algorithm to exhibit data complexity that is independent of the overall database size.  Namely, the algorithm must operate strictly on some aggregated states and the new update batches.  In contrast to online gradient based machine learning algorithms \cite{agarwal2008kernel}, information gain mining can be solved exactly for data streams with strict incremental computation.

In Section~\ref{sec:problemdef}, we formally define the problem of information gain mining in terms of marginal and join entropy measures of attributes of a relation.  We extend the definition to temporal relations with append updates.  In order to focus on efficient online solutions of the problem, we impose a condition on the computational complexity of the incremental algorithm.  In Section~\ref{sec:algorithm}, we construct a solution to the incremental information gain mining problem.  The algorithm maintains a collection of histograms as states.  These states allow the algorithm to compute the {exact information gain} after each updates efficiently.  The algorithm is evaluated in Section~\ref{sec:evaluation}.  We show that the incremental information gain mining can scale to Internet scale temporal relations.

The main contributions of the paper are as follows.

\begin{enumerate}
    \item The formulation of the {\em incremental information gain} problem
    \item An efficient online algorithm that computes the {\em exact information gain} values for temporal relations
    \item A detailed implementation and evaluation of the incremental information gain mining algorithm
\end{enumerate}

\section{Problem Definition}
\label{sec:problemdef}

\begin{figure*}[t]
    \centering
    \includegraphics[width=\textwidth]{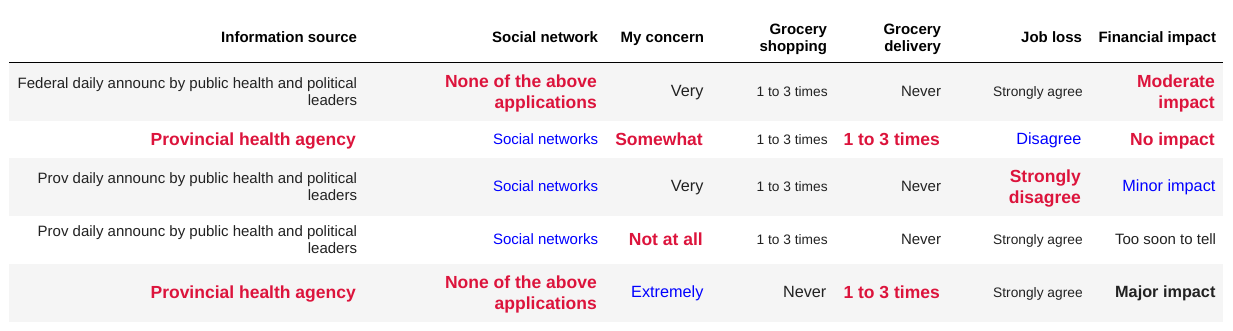}
    \caption{A table view of seven survey answers highlighted according to their information gain with respect to the target attribute \textbf{\em gender}.}
    \label{fig:example_heatmap}
\end{figure*}

\newcommand{\dom}{\mathrm{dom}}

In this section, we will provide the formalism and the problem definition of incremental computation of information gains of relational data streams.

Let $r(A_1, A_2, \dots, A_n)$ be a relation with attributes $\{A_i\}_{1\leq i\leq n}$.  We denote the attributes as $\mathrm{attr}(r)$.  The domain of an attribute is all the distinct values $\dom(A) = \pi_A(r)$ where $\pi_A$ is the relational algebra projection operator.

Given a tuple $t\in r$, we write $t[A]$ to denote the value of $t$ for attribute $A$.
\subsection{Information Gain Mining}
For some relational dataset $r$, we denote the 2-way joint frequency counts as
$$f(A_i = x, A_j = y) = |\{t\in r: t[A_i] = x\ \mathrm{and}\ t[A_j]=y\}|$$

We will write $f_{ij}(x,y) = f(A_i=x, A_j=y)$ whenever there is no confusion over the attributes.

Similarly, the marginal frequency counts are given as:
$$ f_i(x) = f(A_i=x) = |\{t\in r: t[A_i]=x\}|$$

The 2-way joint probabilities are given as
$$p_{i|j}(x, y) = p(A_i=x|A_j=y) = \frac{f_{ij}(x,y)}{f_{j}(y)}$$
where as the marginal probabilities are given as
$$p_i(x) = p(A_i=x) = \frac{f_i(x)}{|r|}$$

We know that an effective way of measuring randomness is the entropy measure of the underlying probability distribution.  So, the {\em randomness} of some attribute $A_i$ can be measured
by the entropy of its marginal probability.
$$H(A_i) = H(p_i) = \sum_{x\in\dom(A_i)} p(A_i=x)\log(p(A_i=x))$$

However, the randomness of the attribute is affected when it is conditioned w.r.t. to another attribute, especially when the conditional attribute is fixed at a particular value.
\begin{eqnarray*}
 && H(A_i|A_j=y) = H(p(A_i|A_j=y)) \\
  &=& \sum_{x\in\dom(A_i)} p(A_i=x|A_j=y)\log(p(A_i=x|A_j=y))
\end{eqnarray*}

Information gain (IG) \cite{cover1991information} is defined as the reduction in entropy.  The condition $A_j = y$ has the information gain given
by:
\begin{equation}
    \mathbf{IG}_i(A_j=y) = I(A_i, A_j=y) = H(A_i)- H(A_i|A_j=y)
    \label{eq:inf-gain}
\end{equation}

By {\em information gain mining}, we refer to the problem identifying pairs $(A_i, y)\in  \mathrm{attr}(r) \times \dom(A_j)$.  We argue that cells with high positive IG are useful to data analysts as they identify situations where the randomness  attribute of interest $A_i$ changes dramatically, thus indicating possible strong dependency (or even causality) relationship between values in $A_i$ and the condition $A_j=y$.

\subsection{Temporal relations and incremental computation}

A relation $r$ is a bag of tuples.  A natural extension of relations over time is {\em temporal relations} \cite{mani2006machine} where $r$ is now a continuous stream of tuples.  In this section, we study the problem of efficient maintenance of $I_i(A_j=y)$ in the presence of incremental updates of the underlying temporal relation.

Consider a single incremental update transaction: $r' = r \cup \Delta r$, where $r$ is the snapshot relation before the update, $\Delta r$ the set of tuples that have been added to the relation, and $r'$ is the final result of the update.

\begin{definition}[Incremental and efficient queries]
Let a query $Q: r\mapsto s$ be a query that derives its result $s$ from a relation $r$.
We say that $Q$ is incremental if there exists some function $H$ such that for all incremental updates $r' = r\cup\Delta r$, we have:
$$ Q(r\cup\Delta r) = H(Q(r), \Delta r)$$
We say that $Q$ is efficient if $H\in\mathcal{O}(\|\Delta r\|)$.
\label{def:incremental}
\end{definition}

Definition\ref{def:incremental} defines {\em view} of temporal relations that can be efficiently maintained over time.  The query $Q$ is the logical definition of the view while the update function $H$ maps the old view to the new view using only the new tuples.  The assertion on the computational complexity of $H$ is to ensure that the view maintenance does not degrade over time as $\|r\|$ grows.

\subsection{Examples}

We illustrate the potential use case of information gain mining with a specific example.  Consider a COVID-19 survey \cite{AB2/SMGRHJ_2020} where individuals are asked to describe how the COVID-19 health crisis impacts their daily lives as well as their reactions to these impacts.

Figure~\ref{fig:example_heatmap} is a view of some subsection of the underlying questionnaire response.  The values are highlighted based on their information gain with respect to the {\em gender} distribution.  The color indicates the degree of information gain: red indicates the 75\%, blue the top 50\%, and black for the values below the 50\% in terms of their information gain.  The information gain values are with respect to the target attribute {\bf Gender}.

\begin{figure*}[t]
    \centering
    \begin{subfigure}[t]{0.3\textwidth}
        \includegraphics[width=\textwidth]{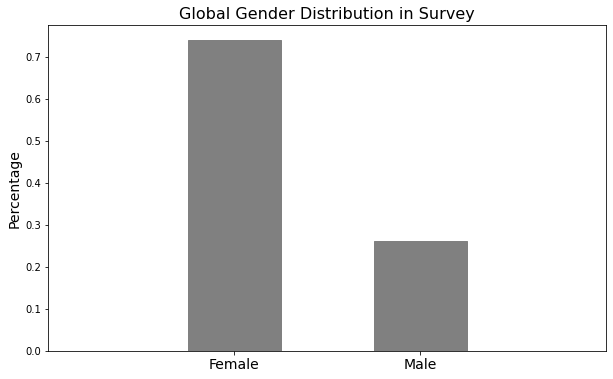}
        \caption{Unconditional distribution of {\em gender}}
    \end{subfigure}
    \begin{subfigure}[t]{0.3\textwidth}
    \includegraphics[width=\textwidth]{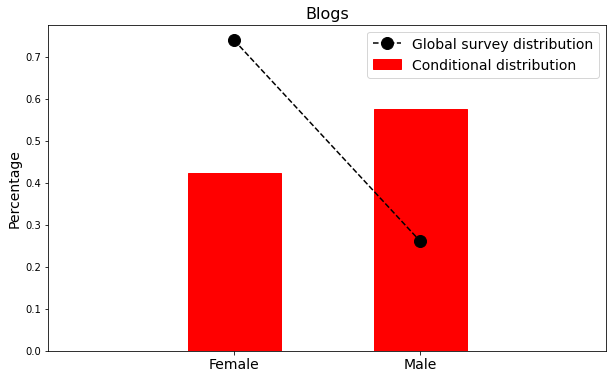}
    \caption{Conditional distribution with high information gain.}
    \end{subfigure}
    \begin{subfigure}[t]{0.3\textwidth}
    \includegraphics[width=\textwidth]{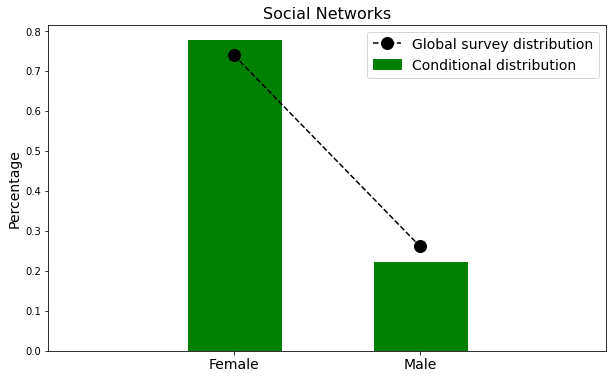}
    \caption{Conditional distribution with low information gain.}
    \end{subfigure}
    \label{fig:dem10}
    \caption{A closer look at the distribution of {\em gender} with respect to different conditions based on information gain mining.}
\end{figure*}

\begin{figure*}[t]
    \centering
    \begin{subfigure}[t]{0.3\textwidth}
        \includegraphics[width=\textwidth]{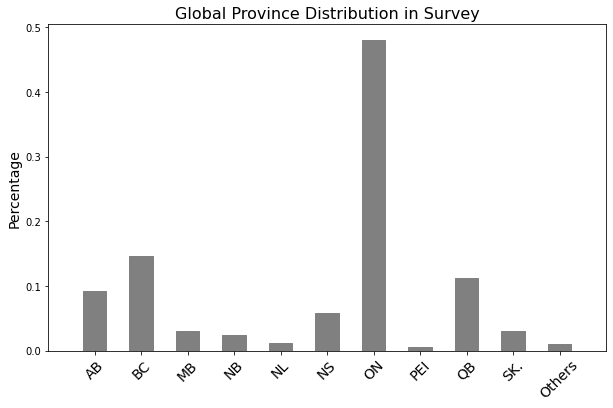}
        \caption{Unconditional distribution of {\em province}}
    \end{subfigure}
    \begin{subfigure}[t]{0.3\textwidth}
    \includegraphics[width=\textwidth]{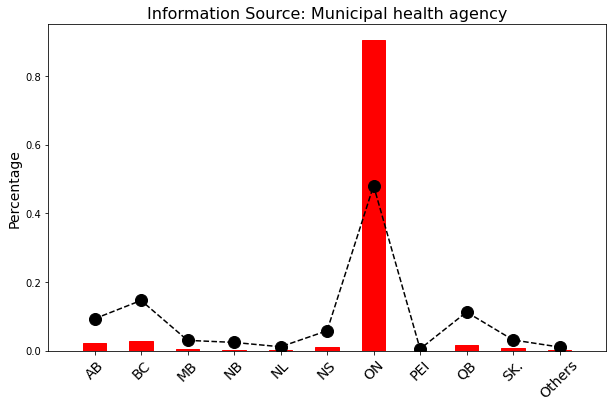}
    \caption{Conditional distribution with high information gain.}
    \end{subfigure}
    \begin{subfigure}[t]{0.3\textwidth}
    \includegraphics[width=\textwidth]{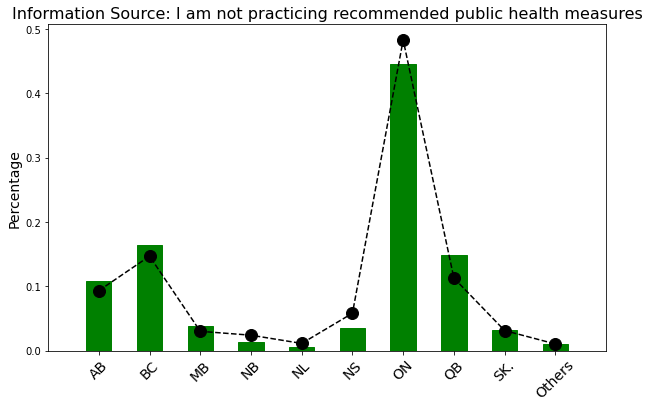}
    \caption{Conditional distribution with low information gain.}
    \end{subfigure}
    \label{fig:prov}
    \caption{A closer look at the distribution of {\em province} with respect to different conditions based on information gain mining.}
\end{figure*}

\section{Algorithm}
\label{sec:algorithm}

\begin{figure}[t]
    \centering
    \begin{tabular}{|l|}\hline
        \textbf{Algorithm:} Initialize States \\
        \textbf{InitStates}($r$) \\ \hline
        for $A_i \in \mathrm{attr}(r)$: \\
        \verb|  | $f_i(x) = |\{t\in r: t[A_i] = x\}|$ \\
        \verb|  | for $A_j \in \mathrm{attr}(r) - \{A_i\}$: \\
        \verb|    | $f_{ij}(x,y) = |\{t\in r: t[A_i]=x\ \mathrm{and}\ t[A_j]=y\}|$ \\
        \verb|  | end for \\
        end for \\
        \hspace{0.45\textwidth} \\
        for $A_i\in \mathrm{attr}(r)$: \\
        \verb|  | $H(A_i)$ = the entropy of $A_i$ \\
        \verb|  | for $A_j\in \mathrm{attr}(r) - \{A_i\}$: \\
        \verb|    | for $y\in\dom(A_j)$: \\
        \verb|      | $H(A_i, A_j=y)$ = conditional entropy of $A_i$ \\
        \verb|    | end for \\
        \verb|  | end for \\
        end for
        \\ \hline
    \end{tabular}
    \label{fig:initstates}
    \caption{A basic algorithm to initialize all the state values.}
\end{figure}

\begin{figure}[t]    
    \begin{tabular}{|l|}\hline
        \textbf{Algorithm:} Incremental Computation of States \\
        \textbf{UpdateStates}($\Delta r$): \\ \hline
        // \textit{Compute $f_{i}$ incrementally} \\
        for $t\in\Delta r$: \\
        \verb|  | for $A_i \in \mathrm{attr}(r)$: \\
        \verb|    | $f_i(t[A_i]) = f_i(t[A_i]) + 1$ \\
        \verb|    | for $A_j\in \mathrm{attr}(r)-\{A_i\}$: \\
        \verb|      | $f_{ij}(t[A_i],t[A_j]) = f_{ij}(t[A_i],t[A_j])+1$\\
        \verb|    | end for \\
        \verb|  | end for \\
        end for \\
        \hspace{0.45\textwidth} \\
        // \textit{Compute post-updated entropy} \\
        for $A_i\in \mathrm{attr}(r)$: \\
        \verb|  | Compute $H'(A_i)$ using Equation~\ref{eq:updated-entropy} and Equation~\ref{eq:incremental-entropy}\\
        \verb|  | for $A_j\in \mathrm{attr}(r)-\{A_i\}$: \\
        \verb|    | for $y\in \dom'(A_j) - \dom(A_j)$: \\
        \verb|      | $H'(A_i|A_j=y) = 0$ \\
        \verb|    | end for \\
        \verb|    | for $y\in\dom(A_j)-\dom'(A_j)$: \\
        \verb|      | $H'(A_i|A_j=y)$ is given by Equation~\ref{eq:incremental-conditional-entropy-1} \\
        \verb|    | end for \\
        \verb|    | for $y\in\dom(A_j)\cap\dom'(A_j)$: \\
        \verb|      | $H'(A_i|A_j=y)$ is given by Equation~\ref{eq:incremental-conditional-entropy-2} \\
        \verb|    | end for \\
        \verb|    | $\mathbf{IG}_i(A_j=y) = H'(A_i) - H'(A_i|A_j=y)$\\
        \verb|  | end for \\
        end for \\ \hline
    \end{tabular}
    
    \label{fig:updatestates}
    \caption{Incremental and efficient computation information gain}
\end{figure}

In this section, we will demonstrate that high information gain mining query is incremental and efficient. We will use the following notations.

\newcommand{\Unchanged}{\mathrm{old}}
\newcommand{\Updated}{\mathrm{updated}}
\newcommand{\New}{\mathrm{new}}

\begin{itemize}
    \item $H(A_i)$ and $H'(A_i)$ are the marginal entropy of $A_i$ before and after the update respectively.
    \item $f_i(x)$ and $f'_i(x)$ are the marginal frequency counts of $A_i=x$ before and after the updates.  We write $\Delta f_i(x)$ as the marginal frequency counts of $A_i=x$ in $\Delta r$.  Thus, we have $f'_i(x) = f_i(x) + \Delta f_i(x)$.
    \item $H(A_i|A_j=y)$ and $H'(A_i| A_j=y)$ are the conditional entropy of $A_i$
    with the condition of $A_j=y$ before and after the update of $\Delta r$.
    \item $f_{ij}(x,y)$ is the joint frequency counts of $A_i=x$ and $A_j=y$.
    \item $\dom'(A_i)$ is the new domain of $A_i$ after update.  We also have
    $\dom'(A_i) = \Unchanged \cup \Updated$, where $\Unchanged \subseteq \dom'(A_i)$ are the values of $A_i$ that are unchanged, $\Updated\subseteq\dom'(A_i)$ are the values of $A_i$ that have their marginal frequencies updated by the insertion of $\Delta r$.
    \item Let $n=\|r\|$ be the tuple count before the update, and $n' = n + \|\Delta r\|$ the tuple count after the update.
\end{itemize}

In the remainder of this section, we will show that both the marginal entropy $H(A_i)$ and the conditional entropy $H(A_i|A_j=y)$ are incremental and efficient queries.

\subsection{Incremental computation of marginal entropy}

First we show that the marginal entropy can be maintained incrementally.  We do so by demonstrating that the post-update marginal entropy $H'(A_i)$ can be computed from the current states involving:

\begin{enumerate}
    \item the marginal frequencies of distinct values: $\{f_{i}(x): x\in\dom(A_i)\}$
    \item the joint frequencies: $\{f_{ij}(x,y): x\in\dom(A_i),\ y\in\dom(A_j)\}$
    \item the marginal entropy: $H_i(A_i)$
\end{enumerate}

Note, it's also crucial that the computation of $H'(A_i)$ is efficient, namely, with complexity in $\mathcal{O}(|\Delta r|)$.

\begin{equation}
\begin{split}
& H'(A_i) \\
&= -\sum_{x\in\dom'(A_i)} p'(A_i)\log p'(A_i) \\
&= -\underbrace{\sum_{x\in\Unchanged} p'(A_i)\log p'(A_i)}_{H_1} \\
    & \quad\quad\quad\quad - \underbrace{\sum_{x\in\Updated\ or\ \New} p'(A_i)\log p'(A_i)}_{H_2}
\label{eq:updated-entropy}
\end{split}
\end{equation}

Note $H'(A_i) \in \mathcal{O}(\|r'\|)$ if we use the naive approach.  Now, we show that it can be done incrementally and efficiently with complexity $\mathcal{O}(\|\Delta r\|)$.
Equation~\ref{eq:updated-entropy} has three components.  We can see that $H_2\in\mathcal{O}(\|\Delta r\|)$ are efficient.  Only the first component  $H_1$ needs to be made incremental.

{\small
\begin{eqnarray*}
H_1 &=& -\sum_{x\in\Unchanged} 
    \left(\frac{f_i(x)}{n'}\right)\log\left(\frac{f_i(x)}{n'}\right) \\
    &=& -\sum_{x\in\Unchanged}
    \frac{n}{n'}\frac{f_i(x)}{n}\cdot\log\left(\frac{f_i(x)}{n}\cdot\frac{n}{n'}\right)\\
    &=& -\frac{n}{n'}
    \sum_{x\in\Unchanged}p_i(x)
    \left[
    \log p_i(x) + \log\left(\frac{n}{n'}\right)
    \right]\\
    &=& -\frac{n}{n'}\sum_{x\in\Unchanged}p_i(x)\log p_i(x)
        -\frac{n}{n'}\log\left(\frac{n}{n'}\right)\sum_{x\in\Unchanged} p_i(x)
\end{eqnarray*}
}

By definition, we have $\dom(A_i) = \Unchanged\ \dot\cup\ \Updated$ (disjoint union).  Hence, we can further simplify the expression for $H_1$ as:

\begin{equation}
\begin{split}
H_1 &=
  -\frac{n}{n'}\left(H(A_i) - \sum_{x\in\Updated}p_i(x)\log p_i(x)\right)\\
  &\quad\quad\quad -\frac{n}{n'}\log\left(\frac{n}{n'}\right)
    \left(1-\sum_{x\in\Updated}p_i(x)\right)
\label{eq:incremental-entropy}
\end{split}
\end{equation}

Equation~\ref{eq:incremental-entropy} is significant in the sense that it computes the entropy component $H_1$ efficiently with computational complexity of $\mathcal{O}(\|\Delta r\|)$.

\subsection{Incremental computation of joint entropy}

Let us now show that the post-update joint entropy, $H'(A_i| A_j=y)$ can also be incrementally and efficiently computed from $H(A_i|A_j=y)$ and the state values.

{\small
\begin{equation}
    \begin{split}
    H'(A_i|A_j=y) &= -\sum_{x\in\dom'(A_i)} p'_{i|j}(x,y)\log p'_{i|j}(x,y) \\
    &= \underbrace{-\sum_{x\in\Unchanged}p'_{i|j}(x,y)\log p'_{i|j}(x,y)}_{H_1(y)} \\
    &\quad\quad \underbrace{-\sum_{x\in\Updated\ \mathrm{or}\ \New}p'_{i|j}(x,y)\log p'_{i|j}(x,y)}_{H_2(y)}
    \end{split}
\end{equation}
}

By similar reasoning as before, $H_2(y)$ is already efficient because it can be done with $\mathcal{O}(\|\Delta r\|)$ time complexity.  We just need to compute $H_1(y)$ incrementally and efficiently.

\begin{eqnarray*}
H_1(y) = -\sum_{x\in\Unchanged}\left(\frac{f'_{ij}(x,y)}{f'_j(y)}\right)
        \log \left(\frac{f'_{ij}(x,y)}{f'_j(y)}\right)
\end{eqnarray*}

We need to consider three cases:

\noindent Case (1): $y$ is new in $\Delta$ and does not appear in $\dom(A_j)$ before the update.  In this case, $f_{ij}(x,y) = 0$, hence $H'_1(y) = 0$.

\vspace{5mm}

\noindent Case (2): $y\in\Unchanged$, namely it does not appear in $\Delta r[A_j]$.  Then,
$f'_{ij}(x,y) = f_{ij}(x,y)$, and $f'_j(y) = f_j(y)$.  Hence,
{\small
\begin{equation}
\begin{split}
& H_1(y) =-\sum_{x\in\Unchanged} p_{i|j}(x,y)\log p_{i|j}(x,y) \\
&= -\left(H(A_i|A_j=y) -\sum_{x\in\Updated} p_{i|j}(x,y)\log p_{i|j}(x,y)\right)
\label{eq:incremental-conditional-entropy-1}
\end{split}
\end{equation}
}

Equation~\ref{eq:incremental-conditional-entropy-1} can be evaluated incrementally and efficiently.

\vspace{5mm}

\noindent Case (3): $y$ appears in both $r$ and $\Delta r$, namely $y\in\Updated$.
Then we have $f'_{ij}(x,y) = f_{ij}(x,y)$ because $x$ is still old, but
$f'_j(y) = f_j(y) + \Delta f_j(y)$.

Thus,
\begin{eqnarray*}
&& p'_{i|j}(x,y) =\frac{p'_{ij}(x,y)}{p'_j(y)} = \frac{f'_{ij}(x,y)}{f'_j(y)} \\
&=& \frac{f_{ij}(x,y)}{f'_j(y)}
   \quad\makebox{because $x$ is old.} \\
&=& \frac{f_{ij}(x,y)}{f_j(y) + \Delta f_j(y)} \\
&=& \frac{f_{ij}(x,y)/f_j(y)}{1 + \Delta f_j(y)/f_j(y)} \\
&=& \frac{p_{i|j}(x,y)}{1 + \Delta f_j(y)/f_j(y)}
\end{eqnarray*}

Substituting this into the expression for $H_1(y)$ from earlier, we get the following.
{\small
\begin{equation}
    \begin{split}
H_1(y) &= -\sum_{x\in\Unchanged}\frac{p_{i|j}(x,y)}{1 + \Delta f_j(y)/f_j(y)}
        \cdot \log \frac{p_{i|j}(x,y)}{1 + \Delta f_j(y)/f_j(y)} \\
&=-\frac{f_j(y)}{f_j(y)+\Delta f_j(y)}
    \Bigg[\sum_{x\in\Unchanged} p_{i|j}(x,y)\log p_{i|j}(x,y) \\
& \quad\quad\quad\quad\quad\quad -\log\left(1+\frac{\Delta f_j(y)}{f_j(y)}\right)\sum_{x\in\Unchanged} p_{i|j}(x,y)\Bigg]
\label{eq:incremental-conditional-entropy-2}        
    \end{split}
\end{equation}
}

We can further decompose the two summation terms in Equation~\ref{eq:incremental-conditional-entropy-2}.

\begin{eqnarray*}
&& \sum_{x\in\Unchanged} p_{i|j}(x,y)\log p_{i|j}(x,y) \\
&=& H(A_i|A_j=y) 
    -\sum_{x\in\Updated} p_{i|j}(x,y)\log p_{i|j}(x,y)
\end{eqnarray*}

And,
\begin{eqnarray*}
&& \sum_{x\in\Unchanged} p_{i|j}(x,y) \\
&=& 1 - \sum_{x\in\Updated} p_{i|j}(x,y)
\end{eqnarray*}

Putting it together, we get the following incremental computation
for $H'(A_i|A_j=y)$.

\section{Evaluation}

\label{sec:evaluation}

\begin{figure}[t]
    \centering
    \begin{subfigure}[t]{0.45\textwidth}
    \includegraphics[width=\textwidth]{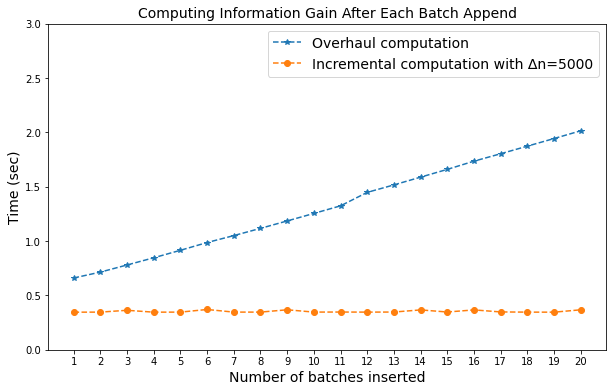}
    \caption{Overhaul vs incremental computation}
    \label{fig:exp-results-overhaul}
    \end{subfigure}
    \begin{subfigure}[t]{0.5\textwidth}
    \includegraphics[width=1.0\textwidth]{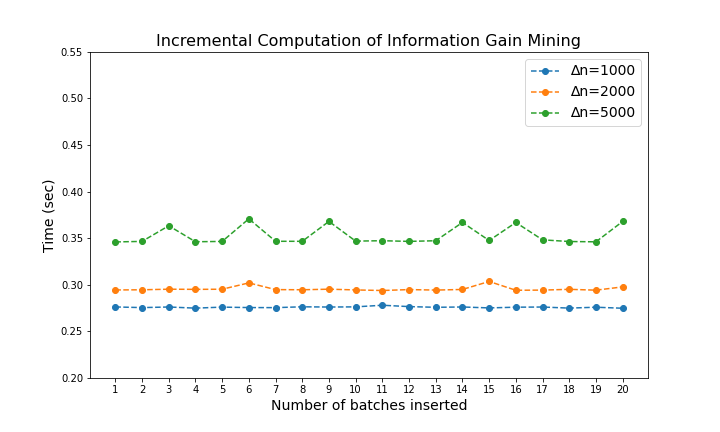}
    \caption{Incremental computation with different batch sizes}
    \label{fig:exp-results-incremental}
    \end{subfigure}
    \caption{Experimental Evaluation}
\end{figure}

\subsection{Implementation}
We implement the proposed algorithm in Python utilizing libraries including Pandas and SciPy. The relational data are stored as Pandas DataFrames. The states of frequencies and entropy are stored as Python dictionaries. We use the Scipy.stats module to compute marginal and conditional entropy. The update rules (Equation~\ref{eq:updated-entropy}, Equation~\ref{eq:incremental-entropy},
Equation~\ref{eq:incremental-conditional-entropy-1} and Equation~\ref{eq:incremental-conditional-entropy-2}) are implemented using standard Pandas DataFrame operators.

\subsection{Experimental Setup}

The experimental setup involves as performing incremental updates of a temporal relation.  The data set used is a publicly available survey response produced and released by Statistics Canada.  We initialized the temporal relation with 1000 tuples, and computed the state variables using {\bf InitStates($r$)}.  Next, we append $\Delta n$ number of tuples incrementally, and performs the updates to the state variables.  The append batch size $\Delta n$ is a parameter we control in the experiment.

We compared our incremental method using {\bf UpdateStates($\Delta r$)} with the naive overhaul method in which {\bf InitStates($r\cup \Delta r$)} is used to compute the new state variables from scratch.

\subsection{Observation}

The overhaul computation approach exhibits linear time complexity, as shown in Figure~\ref{fig:exp-results-overhaul}. This means that the computation of Information Gain will become slower and slower as the temporal relation grows indefinitely. On the other hand, batch based incremental computation has a constant time complexity with respect to the total relation size, but linear with respect to the batch size, as shown in Figure~\ref{fig:exp-results-incremental}. This means that we can continue to track information gain metrics even when the relation grows to unbounded size.

\section{Related Work}
\label{sec:related}

To the authors' knowledge, this paper is the first to provide a formal definition of an algorithmic solution to the efficient incremental information gain mining problem.  Our work is built on a number of previous literature in the area of data integration, information theoretic data mining and incremental computation platforms.

Agarwal et al \cite{agarwal2008kernel} proposed an online support vector machine algorithm which uses an incremental approximation method to compute the support vectors of a kernel based classifier.  Due to the highly non-linear nature of kernel based machine learning, their online computation can only be approximate.

Incremental computation has received great deal of attention from both the database community \cite{schmedding2011incremental} and programming languages \cite{alvarez2019fixing,ryzhyk2019differential}.  Schemdding \cite{schmedding2011incremental} studied the incremental evaluation of database queries expressed as SparQL for linked data in the presence of database updates.  Datalog has been used as a formalism to express the semantics of incremental computation \cite{alvarez2019fixing}, and more recently used as a computing platform to define queries which can be evaluated incrementally \cite{ryzhyk2019differential}.

Information theoretic data mining has been studied for relation databases \cite{mani2006machine,luo2018scalable,lippi2011relational,Nowozin12improvedinformation}.  Information gain has already been shown to be an useful measure for relational variable assessment \cite{lippi2011relational} and improvements in decision tree construction \cite{Nowozin12improvedinformation}.  Luo et al \cite{luo2018scalable} studied scalable linear algebraic operations that can be applied to relational databases.

\section{Conclusion and Future Work}

\subsection{Conclusion}
In this paper, we have proposed the information gain mining problem.  We formally defined the problem of computing the information gain of relational values with respect to selected target attributes.  The usefulness of Information Gain Mining is illustrated with case studies based on actual public survey data sets from Statistics Canada.  We can demonstrate that the most salient relational values obtained from the mining algorithm reveal dependencies and insights embedded in the relational data set.  We generalized the information gain mining algorithm to temporal relations where new batches of tuples are appended continuously to the relational data set.  We analytically showed that the information gain mining of temporal relations can be efficiently and incrementally computed.  By defining the state variables as simple frequency counters and entropy measures, we demonstrate that the states can be maintained incrementally and the information gain computed efficiently from these state variables.

Our incremental information gain algorithm is evaluated and demonstrated constant time computational complexity with respect to the growing temporal relation size.

\subsection{Future Work}
We plan to extend this work in several directions.  Currently, we do not have any discriminatory selection over the target attributes $A_i$ and the conditional relational value $A_j=y$ when computing the information gain measure $\mathbf{IG}_i(A_j=z)$.  The current implementation computes the information gain with respect to all target attributes $A_i\in \mathrm{attr}(r)$.  As future work, we play to utilize tools from linear algebra and statistics to identify uncorrelated attributes, and thus potentially significantly pruning the computation space.

\bibliographystyle{plain}
\bibliography{my/references.bib}

\end{document}